# Fault prediction in aircraft engines using Self-Organizing Maps

Marie Cottrell[1], Patrice Gaubert[1], Cédric Eloy[2], Damien François[2],
Geoffroy Hallaux[2], Jérôme Lacaille[3], Michel Verleysen[2]

[1] SAMOS-MATISSE, UMR CNRS CES, Université Paris 1 Panthéon Sorbonne, France
[2] Machine Learning Group, Université catholique de Louvain, Belgium
[3] SNECMA YYE, Villaroche, France
Marie.cottrell@univ-paris1.fr

**Abstract.** Aircraft engines are designed to be used during several tens of years. Their maintenance is a challenging and costly task, for obvious security reasons. The goal is to ensure a proper operation of the engines, in all conditions, with a zero probability of failure, while taking into account aging. The fact that the same engine is sometimes used on several aircrafts has to be taken into account too.

The maintenance can be improved if an efficient procedure for the prediction of failures is implemented. The primary source of information on the health of the engines comes from measurement during flights. Several variables such as the core speed, the oil pressure and quantity, the fan speed, etc. are measured, together with environmental variables such as the outside temperature, altitude, aircraft speed, etc.

In this paper, we describe the design of a procedure aiming at visualizing successive data measured on aircraft engines. The data are multi-dimensional measurements on the engines, which are projected on a self-organizing map in order to allow us to follow the trajectories of these data over time. The trajectories consist in a succession of points on the map, each of them corresponding to the two-dimensional projection of the multi-dimensional vector of engine measurements. Analyzing the trajectories aims at visualizing any deviation from a normal behavior, making it possible to anticipate an operation failure.

However rough engine measurements are inappropriate for such an analysis; they are indeed influenced by external conditions, and may in addition vary between engines. In this work, we first process the data by a General Linear Model (GLM), to eliminate the effect of engines and of measured environmental conditions. The residuals are then used as inputs to a Self-Organizing Map for the easy visualization of trajectories.

**Keywords:** aircraft engine maintenance, fault detection, general linear models, self-organizing maps.

# 1 Introduction

Security issues in the aircrafts are a major concern for obvious reasons. Among the many aspects of security issues, ensuring a proper operation of engines over their lifetime is an important task.

Aircraft engines are built with a high level of security norms. They undergo regularly a full maintenance with disassembling, replacement of parts, etc. In addition, between two such maintenances, many parameters are measured on the engines during the flights. These parameters are recorded, and used both at short and long terms for immediate action and alarm generation respectively.

In this work, we are interested in the long-term monitoring of aircraft engines. Measurements on the engines during flights are used to detect any deviation from a "normal" behavior, making it possible to anticipate possible faults. This fault anticipation is aimed to facilitate the maintenance of aircraft engines.

Self-Organizing Maps are here used to provide experts a supplementary tool to visualize easily the evolution of the data measured on the engines. The evolution is characterized by a trajectory on the two-dimensional Self-Organizing Map. Abnormal aging and fault appearance will result in deviation of this trajectory, with respect to normal conditions. The output of this data mining study is therefore a visual tool that can be used by experts, in addition to their traditional tools based on quantitative inspection of some measured variables. Self-Organizing Maps are useful tools for fault detection and prediction in plants and machines (see [1], [2], [3], [4], [5], for example).

Analyzing the rough variables measured on the engines during flights is however not appropriate. Indeed these measurements may vary from one engine to another, and may also vary according to "environmental" conditions (such as the altitude, the outside temperature, the speed of the aircraft, etc.). In this work, we first remove the effects of environmental (measured) variables, and the engine effects, from the rough measurements. The residuals of the regression are then used for further analysis by Self-Organizing maps.

The following of this paper is organized as follows. In Section 2, the data are described and notations are defined. Section 3 presents the methodology: Section 3.1 describes how the effects of engines and of environmental variables are removed by a General Linear Model, and Section 3.2 shows the visual analysis of the GLM residuals by Self-Organizing Maps. Section 4 describes the experimental results, before some conclusions in Section 5.

# 2 Data

Measurements are collected on a set of $I$ engines. On each engine $i$ ($1 \leq i \leq I$), $n_i$ sets of measurements are performed successively. Usually one set is measured during each flight; there is thus no guarantee that the time intervals between two sets of measures are approximately equal. Each set of observations is denoted by $Z_{ij}$, with $1 \leq i \leq I$ and $1 \leq j \leq n_i$.

Each set $Z_{ij}$ contains both variables related to the behavior of the engine, and variables that are related to the environment. Let us denote the $p$ engine variables by $Y_{ij}^1,..., Y_{ij}^p$ and the $q$ environmental variables by $X_{ij}^1,..., X_{ij}^q$. Each set of measurements is thus a vector $Z_{ij}$, where

$$Z_{ij} = (Y_{ij}, X_{ij}) = (Y_{ij}^1,..., Y_{ij}^p, X_{ij}^1, ..., X_{ij}^q) \ . \qquad (1)$$

In this study, the variables at disposal are those listed in Table 1.

**Table 1.** Engine and environmental variables

| Engine variables | | Environmental variables | |
|---|---|---|---|
| $Y_{ij}^1$ | core speed | $X_{ij}^1$ | Mach |
| $Y_{ij}^2$ | oil pressure | $X_{ij}^2$ | Engine bleed valve 1 |
| $Y_{ij}^3$ | HPC discharge stat. pres. | $X_{ij}^3$ | Engine bleed valve 2 |
| $Y_{ij}^4$ | HPC discharge temp. | $X_{ij}^4$ | Engine bleed valve 3 |
| $Y_{ij}^5$ | Exhaust gas temp. | $X_{ij}^5$ | Engine bleed valve 4 |
| $Y_{ij}^6$ | Oil temperature | $X_{ij}^6$ | Isolation valve left |
| $Y_{ij}^7$ | Fuel flow | $X_{ij}^7$ | Altitude |
| | | $X_{ij}^8$ | HPT active clearance |
| | | $X_{ij}^9$ | LPT active clearance |
| | | $X_{ij}^{10}$ | Total air temperature |
| | | $X_{ij}^{11}$ | Nacelle temperature |
| | | $X_{ij}^{12}$ | ECS Pack 1 flow |
| | | $X_{ij}^{13}$ | ECS Pack 2 flow |

The goal of this study is to visualize the $Y_{ij}$ vectors. The visualization of the successive measurements $j$ for a specific engine $i$ corresponds to a trajectory.

## 3 Methodology

Rough $Y_{ij}$ measurements of the engine variables cannot be used as such for the analysis. Indeed the $Y_{ij}$ strongly depend on

- engine effects, i.e. the fact that the engines may differ, and on
- environmental effects, i.e. the dependence of the engine variables $Y_{ij}$ on the environmental conditions $X_{ij}$.

Both dependences lead to differences in observed variables that have nothing to do with aging or fault anticipation. It is therefore important to remove these effects before further analysis.

In this work, we use a GLM (General Linear Model) [6] to remove these effects, since the independent variables are of two types : categorical (engine effect) and real-valued (environmental variables). The use of GLM implies two hypotheses. First, it is assumed that the effect of the environment is effectively measured in the environmental variables $X_{ij}$; obviously, non-measured effects cannot be removed. Secondly, it is assumed that the relation between the engine variables $Y_{ij}$ and the environmental variables $X_{ij}$ is linear; this last assumption is probably not perfectly correct, but it will be shown in the experimental section that even under this hypothesis, the statistical significance of the $X_{ij}$ effects is high; this justifies a posteriori to remove at least the linear part (first-order approximation) of the relation.

The residuals of the regression of the $Y_{ij}$ variables over the $X_{ij}$ ones and the motor effects are then used for the analysis. A Self-Organizing Map is used to visualize the two-dimensional projection of the residuals corresponding to each vector $Y_{ij}$. Then, the different states $j$ ($1 \leq j \leq n_i$) of a single engine are linked together to form a trajectory.

The next two subsections detail how to perform the GLM regression on the engine variables, and how to use the Self-Organizing Maps on the GLM residuals.

### 3.1 Computation of the residuals (so-called corrected values)

The computation of the values obtained by removing the effects of the environment variables and of the engine is done by using a General Linear Model, where the explanatory variables are of two kinds: one variable is categorical (the engine number), the others are real-valued variables (the environment variables).

For each engine variable $m = 1, \ldots, p$, the GLM model can be written as:

$$Y_{ij}^m = \mu^m + \alpha_i^m + \lambda_1^m X_{ij}^1 + \ldots + \lambda_q^m X_{ij}^q + \varepsilon_{ij}^m, \quad (2)$$

where $i = 1, \ldots, I$, is the engine number, $j$ is the flight number, $\alpha_i^m$ is the engine effect on the $m$-th variable, $X_{ij}^1, \ldots, X_{ij}^q$ are the environmental variables, $\lambda_1^m, \ldots, \lambda_q^m$ are the regression coefficients for the $m$-th variable, and the error term $\varepsilon_{ij}^m$ is centered with variance $\sigma_m^2$. The parameters $\alpha_i^m, \lambda_1^m, \ldots, \lambda_q^m$, are estimated by the least squares method, and in order to avoid colinearity, we have to add the constraint $\sum_{i=1}^{I} n_i \alpha_i^m = 0$.

Note that it is possible to model the motor effect by a random term $A_i^m$ instead of the fixed effect $\alpha_i^m$; $A_i^m$ is also supposed to be centered with variance $\sigma_A^2$. Even if the model is slightly different, the residuals are the same.

Fisher statistics allows us to verify the significance of the models and to confirm the interest of the adjustment of engine variable for the environmental ones and the motor effect.

Let us denote by $R_{ij}^m$, $m = 1, \ldots, p$ the residuals (2), equal to the estimated values $\hat{\varepsilon}_{ij}^m$. The residuals are the values adjusted for the motor effect and the environment variables.

### 3.2 Self-Organizing Maps on the residuals

Next we consider a $n$ by $n$ Kohonen map [7] and train it with the $p$-dimensional residuals $R_{ij}^m$ ($m = 1, \ldots, p$). We use the SOM toolbox for Matlab [8] for the experiments. In that way, each flight $j$ of each engine $i$ is projected on a Kohonen class on the map. We can identify the different locations on the map by looking at the corresponding code-vectors and at their components, and then give a description of the clusters. For each engine $i$, we define the sequence of the class numbers corresponding to the successive flights $j = 1, \ldots, n_i$. This sequence is the trajectory of engine $i$. In this way we get a visual representation of the successive states of the engines on the Kohonen map. Then we can compare these trajectories by introducing a measure of distance between them.

## 4 Experiments

We consider real data which consist in the observation of $I = 91$ engines. Each engine is measured for a number of flights between 500 and 800. There are 7 engine variables and 13 environment variables, as illustrated in Table 1.

### 4.1 Justification of the computation of adjusted variables

To justify the computation of the residuals (i.e. the values adjusted for engine effect and environment variables), we can for example show the result of a PCA on the raw data and use different colors for 5 different engines. We see (Fig.1, left) that each engine clearly defines a cluster in the projection on the first two principal components. Fig.1, right also shows that the histograms of the engine variables ($Y_{ij}4$ is illustrated) depend on the engine.

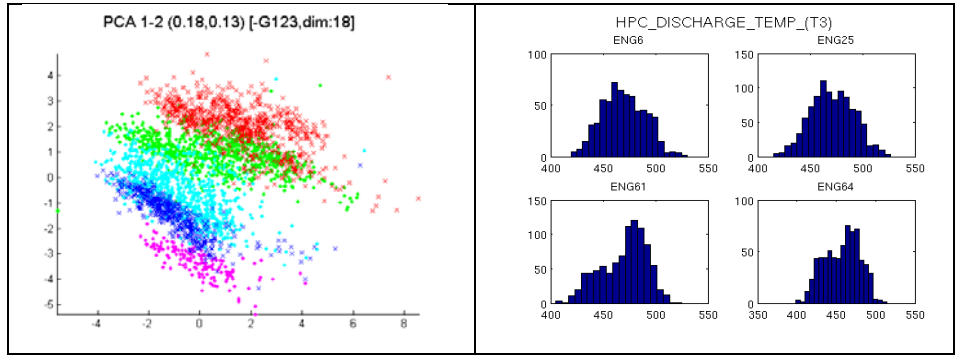

**Fig. 1.** Left: the first two principal components for five engines. The data are the 7-dimensional engine variables. Right; the values of variable $Y^4$ (HPC discharge temperature) for 4 engines.

The correlation between variables can be illustrated too. As an example, Figure 2 shows variable $Y^5$ (EGT) as a function of variable $X^{10}$ (Total Air Temperature) in four engines. It is obvious that both variables are strongly dependent.

These few examples clearly show that it is necessary to remove the effects of the engine and of the environmental variables, by computing the residuals in model (2).

### 4.2 Self-Organizing Maps on adjusted variables and trajectories

After the extraction of the residuals $R_{ij}$ as detailed in Section 3.1, the second step consists in training a 20 by 20 Kohonen map on these residuals. Figure 3 shows the map obtained, colored according to each of the 7 engine variables. It is clearly visible that the organization of the map is successful (all variables are smoothly varying on the map).

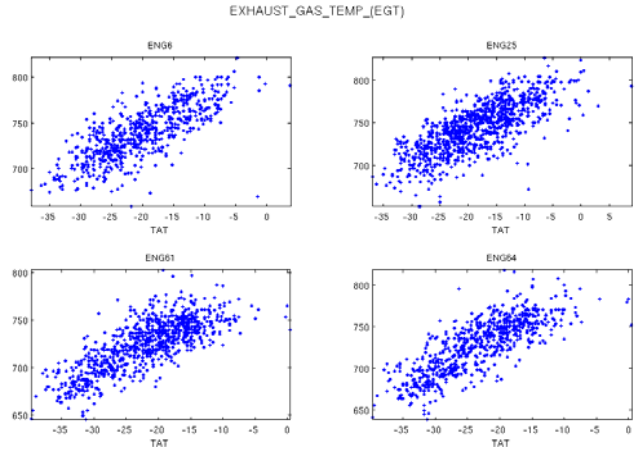

**Fig. 2.** Almost linear dependence between variable $Y^5$ (EGT) and variable $X^{10}$ (Total Air Temperature).

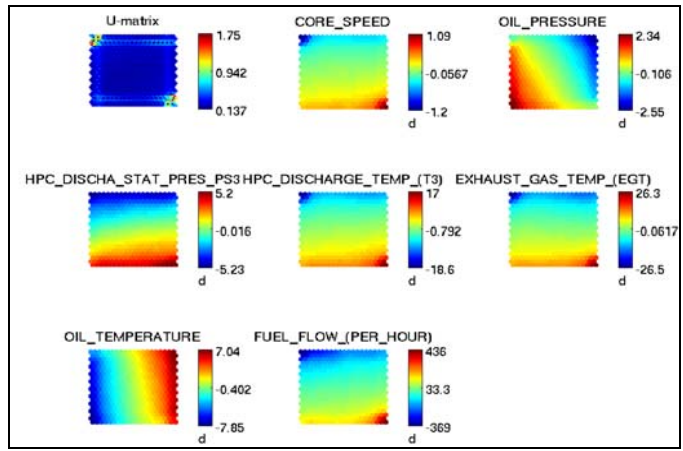

**Fig. 3.** 20x20 self-organizing map on the residuals. The first plot shows the U-matrix, the other ones display the distribution of the 7 engine variables $R^1 – R^7$ over the map.

We can see that variables $R^1$, $R^3$, $R^4$, $R^5$, $R^7$ on one hand, and $R^2$ and $R^6$ on another hand, form high-correlated groups of variables (his property can be verified by computing the correlation matrix).

The 400 classes are then grouped (hierarchical clustering) into 5 super-classes, as shown in Figure 4. Finally, Figure 5 shows the trajectories of the engines. As examples, the trajectories of engines 6, 25 and 88 are illustrated.

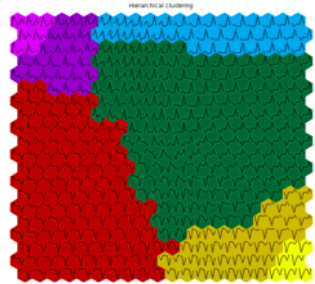

**Fig. 4.** Five super-classes are shown after hierarchical clustering of the 400 classes. The centroids are also shown inside each class.

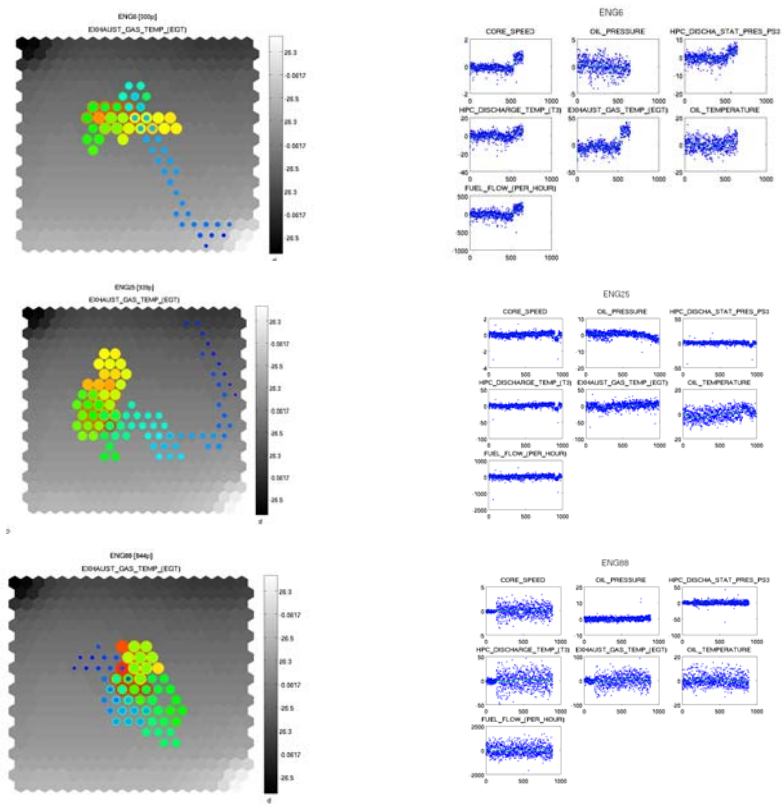

**Fig. 5.** Left, trajectories of engines 6, 25 and 88 on the Kohonen map; the dots color indicates the evolution along the trajectory (from red to blue, through yellow and green). The background shows the level of the EGT variable ($R^5$). Right: the residuals for the same engines.

We observe that the trajectories have different shapes. Looking at the graphs of the 7 adjusted engine variables (Figure 5 right), we conclude that the visual representations on the Kohonen map provide a synthetic representation for the temporal evolution of the engines.

The next step is then to characterize the different shapes of trajectories, to define a suitable distance measure between these trajectories, and to define typical behaviors related to typical faults.

## 5  Conclusions

The proposed method is a useful tool to summarize and represent the temporal evolution of an aircraft engine flight after flight. Further work will consist in defining classes for the trajectories and in associating each class to some specific behavior. Using the maintenance reports which contain the a posteriori measured data related to each engine, it will be possible to identify the classes with possible failures. So the visual examination of such trajectories will help anticipating faults in aircraft engines.

## References

1. Goser K., Metzen S., Tryba V. "Designing of basic Integrated Circuits by Self-organizing Feature Maps, Neuro-Nîmes, 1989
2  Barreto G.A., Mota J.C..M., Souza L.G.M., Frota R.A., Aguayo L. "Condition monitoring of 3G cellular networks through competitive neural models", IEEE Transactions on Neural Networks, vol. 16, no. 5, pp. 1064-1075, 2005.
3  Sarasamma S.T., Zhu Q.A. "Min-max hyperellipsoidal clustering for anomaly detection in network security", IEEE Transactions on Systems, Man and Cybernetics, Part B, vol. 36, no. 4, pp. 887-901, 2006.
4  Svensson, M., Byttner, S., & Rögnvaldsson, T. "Self-organizing maps for automatic fault detection in a vehicle cooling system", Intelligent Systems, 2008. IS '08. 4th International IEEE Conference, vol. 3, pp. 24-8-24-12, 6-8 Sept. 2008. DOI: 10.1109/IS.2008.4670481
5  Alhoniemi E., Simula O., Vesanto J. "Process monotoring and modeling using the self-organizing", Integrated Computer Aided Engineering, vol. 6, no. 1, pp. 3-14, 1999.
6. Draper, N.R., Smith H.: Applied Regression Analysis. John Wiley & Sons, New York (1966)
7. Kohonen, T.: Self-Organizing Maps, Springer Series in Information Sciences, Vol 30, Springer (1995).
8  Laboratory of Computer and Information Science, Helsinky University of Technology, SOM Toolbox for Matlab, *www.cis.hut.fi/projects/**somtoolbox***